\documentclass[sigconf]{acmart}
\usepackage{mathtools}
\usepackage{subcaption}
\usepackage{float}
\usepackage{stfloats}
\usepackage{makecell}
\usepackage{multirow}
\usepackage{enumitem}

\AtBeginDocument{%
  }

\setcopyright{acmlicensed}
\copyrightyear{2025}
\acmYear{2025}
\acmDOI{XXXXXXX.XXXXXXX}
\acmConference[Conference acronym 'XX]{Make sure to enter the correct
  conference title from your rights confirmation email}{June 03--05,
  2025}{Woodstock, NY}

\begin{document}

\title[HyFormer]{HyFormer: Revisiting the Roles of Sequence Modeling and Feature Interaction in CTR Prediction}

\author{Yunwen Huang*}
\thanks{*These authors contributed equally.}
\thanks{\textsuperscript{†}Corresponding Authors.}
\email{huangyunwen.eleanor@bytedance.com}
\affiliation{%
  \institution{ByteDance AML}
  \city{Beijing}
  \country{China}
}

\author{Shiyong Hong*}
\email{hongshiyong.66@bytedance.com}
\affiliation{%
  \institution{ByteDance Search}
  \city{Beijing}
  \country{China}
}

\author{Xijun Xiao*}
\email{xiaoxijun@bytedance.com}
\affiliation{%
  \institution{ByteDance AML}
  \city{Beijing}
  \country{China}
}

\author{Jinqiu Jin*}
\email{jinjinqiu.02@bytedance.com}
\affiliation{%
  \institution{ByteDance Search}
  \city{Beijing}
  \country{China}
}

\author{Xuanyuan Luo}
\email{xuanyuanluo@bytedance.com}
\affiliation{%
  \institution{ByteDance AML}
  \city{Hangzhou}
  \country{China}
}

\author{Zhe Wang}
\email{zhewang.tim@gmail.com}
\affiliation{%
  \institution{ByteDance Search}
  \city{Beijing}
  \country{China}
}

\author{Zheng Chai\textsuperscript{†}}
\email{chaizheng.cz@bytedance.com}
\affiliation{%
  \institution{ByteDance AML}
  \city{Hangzhou}
  \country{China}
}

\author{Shikang Wu\textsuperscript{†}}
\email{wushikang@bytedance.com}
\affiliation{%
  \institution{ByteDance Search}
  \city{Beijing}
  \country{China}
}

\author{Yuchao Zheng}
\email{zhengyuchao.yc@bytedance.com}
\affiliation{%
  \institution{ByteDance AML}
  \city{Hangzhou}
  \country{China}
}

\author{Jingjian Lin}
\email{linjingjian000@gmail.com}
\affiliation{%
  \institution{ByteDance Search}
  \city{Beijing}
  \country{China}
}
\renewcommand{\shortauthors}{Huang et al.}

\begin{abstract}
Industrial large-scale recommendation models (LRMs) face the challenge of jointly modeling long-range user behavior sequences and heterogeneous non-sequential features under strict efficiency constraints. However, most existing architectures employ a decoupled pipeline: long sequences are first compressed with a query-token based sequence compressor like LONGER, followed by fusion with dense features through token-mixing modules like RankMixer, which thereby limits both the representation capacity and the interaction flexibility. This paper presents \textbf{HyFormer}, a unified hybrid transformer architecture that tightly integrates long-sequence modeling and feature interaction into a single backbone. From the perspective of sequence modeling, we revisit and redesign query tokens in LRMs, and frame the LRM modeling task as an alternating optimization process that integrates two core components: \emph{Query Decoding} which expands non-sequential features into \emph{Global Tokens} and performs long sequence decoding over layer-wise key-value representations of long behavioral sequences; and \emph{Query Boosting} which enhances cross-query and cross-sequence heterogeneous interactions via efficient token mixing. The two complementary mechanisms are performed iteratively to refine semantic representations across layers. Extensive experiments on billion-scale industrial datasets demonstrate that HyFormer consistently outperforms strong LONGER and RankMixer baselines under comparable parameter and FLOPs budgets, while exhibiting superior scaling behavior with increasing parameters and FLOPs. Large-scale online A/B tests in high-traffic production systems further validate its effectiveness, showing significant gains over deployed state-of-the-art models. These results highlight the practicality and scalability of HyFormer as a unified modeling framework for industrial LRMs.
\end{abstract}





\maketitle

\section{Introduction}

Modern industrial large-scale recommendation models (LRMs) operate in increasingly complex environments, where accurate prediction relies on jointly modeling long-range user behavior histories and rich heterogeneous features, including user profiles, contextual signals, and cross features. As user engagement grows over extended time horizons and feature spaces continue to expand, effectively integrating long sequential signals with high-dimensional non-sequential information has become a central challenge for large-scale recommendation and search systems.
To address this challenge, recent industrial architectures have largely converged on a separated scaling paradigm that combines \emph{long sequence modeling}\cite{Xu_2025,Zivic_2024,borisyuk2024lirank} with \emph{feature interaction} \cite{gui2023hiformer,yu2025hhfthierarchicalheterogeneousfeature,xu2025store,khrylchenko2025scaling}. Within this paradigm, long user behavior sequences are encoded by dedicated sequence transformers to capture temporal dependencies and user interests, and the compressed sequence token(s) are mixed with other heterogeneous features through token-mixing or interaction modules to enable cross-feature reasoning. This ``Long Sequence Modeling, Then Heterogeneous Feature Interaction'' pipeline has proven effective and has become the dominant design choice for scaling up modern industrial LRMs. Despite strong empirical performance, this prevailing paradigm fundamentally enforces a compressed, late-fusion, and unidirectional interaction pattern. As sequence lengths and model capacities continue to increase, this two-stage design reveals fundamental limitations that restrict both modeling expressiveness and scalability.

\begin{itemize}[leftmargin=*]
    \item Sequence transformers in existing architectures often rely on overly simplified query representations \cite{zhou2018deep,zhou2019deep,feng2019deep} during sequence compression. In practice, the query tokens used to aggregate long behavior sequences are usually derived from a limited subset of candidate-related or global features, constraining the amount of contextual information that can be leveraged when modeling long-term user interests. However, directly increasing the number of query tokens would lead to a significant degradation in serving efficiency under KV-Cache and M-Falcon mechanisms \cite{zhai2024actions, chai2025longerscalinglongsequence}.
    \item Interactions between sequence-compressed tokens and heterogeneous non-sequential tokens typically occur only at late stages of the model. Under the current paradigm, cross-feature reasoning is deferred until after sequence compression, leading to shallow and implicit interactions between different token types. This delayed fusion limits the model's ability to capture fine-grained dependencies across multiple behavior sequences and heterogeneous feature groups, and prevents early-layer representations from benefiting from cross-domain contextual information.
    \item Since interaction modules operate only on compressed sequence representations, increasing model capacity or sequence length primarily improves isolated components rather than enhancing joint representations. As a result, scaling up depth or parameters leads to a lower scaling efficiency, where performance improvements increase at a slower rate with respect to additional computational budgets, as computation is less effectively translated into richer joint representations. 

\end{itemize}


These limitations motivate a fundamental rethinking of how long-range sequence modeling and heterogeneous feature interaction should be integrated. Rather than treating sequence encoding and token mixing as two loosely coupled stages, a unified modeling framework is needed to enable deeper, earlier, and bidirectional interactions between sequential and non-sequential signals.

In this paper, we propose \textbf{HyFormer}, a \emph{hybrid transformer} architecture that unifies sequence modeling and feature interaction within a single backbone. HyFormer introduces a set of \textbf{global tokens} that serve as a shared semantic interface between long behavior sequences and heterogeneous features. Through a stacked design, HyFormer alternates between two lightweight yet expressive mechanisms. The \emph{Query Decoding} module uses global query tokens to attend over layer-wise key--value representations of long behavioral sequences, allowing global context to directly shape sequence representations. The \emph{Query Boosting} module further strengthens cross-query and cross-sequence interactions via efficient token mixing, progressively enriching semantic representations across layers. This design enables a bidirectional flow of information between sequence modeling and feature interaction components, overcoming the limitations of traditional decoupled pipelines. Extensive experiments on billion-scale industrial datasets demonstrate that HyFormer consistently outperforms strong sequence-based and token-mixing baselines under comparable parameter and FLOPs budgets. Moreover, HyFormer exhibits superior scaling behavior with respect to model FLOPs and parameters, and achieves significant gains in large-scale online A/B tests deployed in high-traffic production systems.

In summary, this paper makes the following contributions:
\begin{itemize}
    \item We identify fundamental limitations of the prevailing decoupled sequence modeling and feature interaction paradigm in large-scale industrial recommender systems, and analyze how its unidirectional and late-fusion design constrains modeling capacity and scalability.
    \item We propose HyFormer, a unified hybrid transformer architecture that enables bidirectional, layer-wise interaction between long-range behavioral sequences and heterogeneous features through Query Decoding and Query Boosting, achieving state-of-the-art performance and scalability in real-world industrial settings.
    \item We empirically verify the effectiveness and its superior scaling performance of the proposed models on a billion-scaled industrial dataset. Currently, HyFormer has been fully deployed at Bytedance, serving billions of users each day.
\end{itemize}

\section{Related Work}

\subsection{Traditional Recommendation Paradigms}

Modern industrial LRMs are typically built upon two major components: behavior-sequence modeling and feature-interaction networks. In this paradigm, user behavior histories are first encoded by dedicated sequence models, whose outputs are then consumed by downstream interaction modules together with heterogeneous non-sequential features. Recent industrial systems have substantially advanced the scalability of sequence modeling along this direction. Methods such as SIM~\cite{qi2020searchbasedusermodelinglifelong}, ETA~\cite{chen2022efficient},TWIN~\cite{chang2023twintwostagenetworklifelong,Si_2024}, TransAct~\cite{Xia_2023}, and LONGER~\cite{chai2025longerscalinglongsequence} extend sequence encoders to hundreds or thousands of events through efficient attention mechanisms, hierarchical aggregation, KV caching, and serving-friendly designs. These works demonstrate clear power-law scaling trends in modeling long-range user behaviors under large-scale traffic, while largely preserving a two-stage architecture that decouples sequence encoding from feature interaction.

On the feature-interaction side, early models such as DeepFM\cite{guo2017deepfmfactorizationmachinebasedneural}, xDeepFM\cite{Lian_2018}, and DCNv2\cite{Wang_2021} model low-order or bounded-degree feature crosses at scale but suffer from diminishing returns as interaction depth increases.  Recent scaling studies like Wukong\cite{zhang2024wukong} and RankMixer\cite{zhu2025rankmixerscalingrankingmodels} highlight that cross-module expansion becomes a key driver of industrial performance. These models represent the current state of large-scale feature-interaction design, yet the interaction stack and sequence encoder remain loosely coupled in most production pipelines, resulting in late fusion and preventing unified optimization across heterogeneous signals.

 
\subsection{Unified Recommendation Architectures}

To reduce the fragmentation between sequence modeling and feature interaction, recent studies explore unified architectures that handle heterogeneous signals within a single backbone. Hierarchical generative architectures such as HSTU\cite{zhai2024actionsspeaklouderwords} represent a unified
recommendation paradigm by performing sequence transduction conditioned on contextual and candidate signals. InterFormer\cite{zeng2025interformereffectiveheterogeneousinteraction} bridges the gap between sequence encoders and interaction networks by introducing learnable interaction tokens that enable bidirectional signal exchang. MTGR \cite{Han_2025} further pushes unification by reorganizing user, behavior, real-time, and candidate features into heterogeneous tokens and encoding them with a shared Transformer-style backbone, enabling both sequence information and cross features to be modeled coherently. Following MTGR, OneTrans\cite{zhang2025onetransunifiedfeatureinteraction} shares a similar direction by using a single Transformer to jointly capture sequence dependencies and high-order feature interaction, while simplifying the Transformer structure with a pyramid-compression style. This work can be regarded as a simplified version compared to MTGR. 

As MTGR\cite{Han_2025} and OneTrans\cite{zhang2025onetransunifiedfeatureinteraction} simply increase the number of query tokens as the number of all the non-sequence tokens, it would be readily observed a significant drop in serving efficiency in practice (see Section 4). Besides, a unified transformer structure for modeling feature interaction is generally insufficient in industrial-scale LRMs \cite{zhu2025rankmixerscalingrankingmodels}. Overall, unified architectures represent a step toward dissolving the long-standing separation between sequence models and feature-interaction stacks, though achieving full unification with minimal architectural overhead remains an open challenge.


\begin{figure*}[t]
    \centering
    \includegraphics[width=\textwidth]{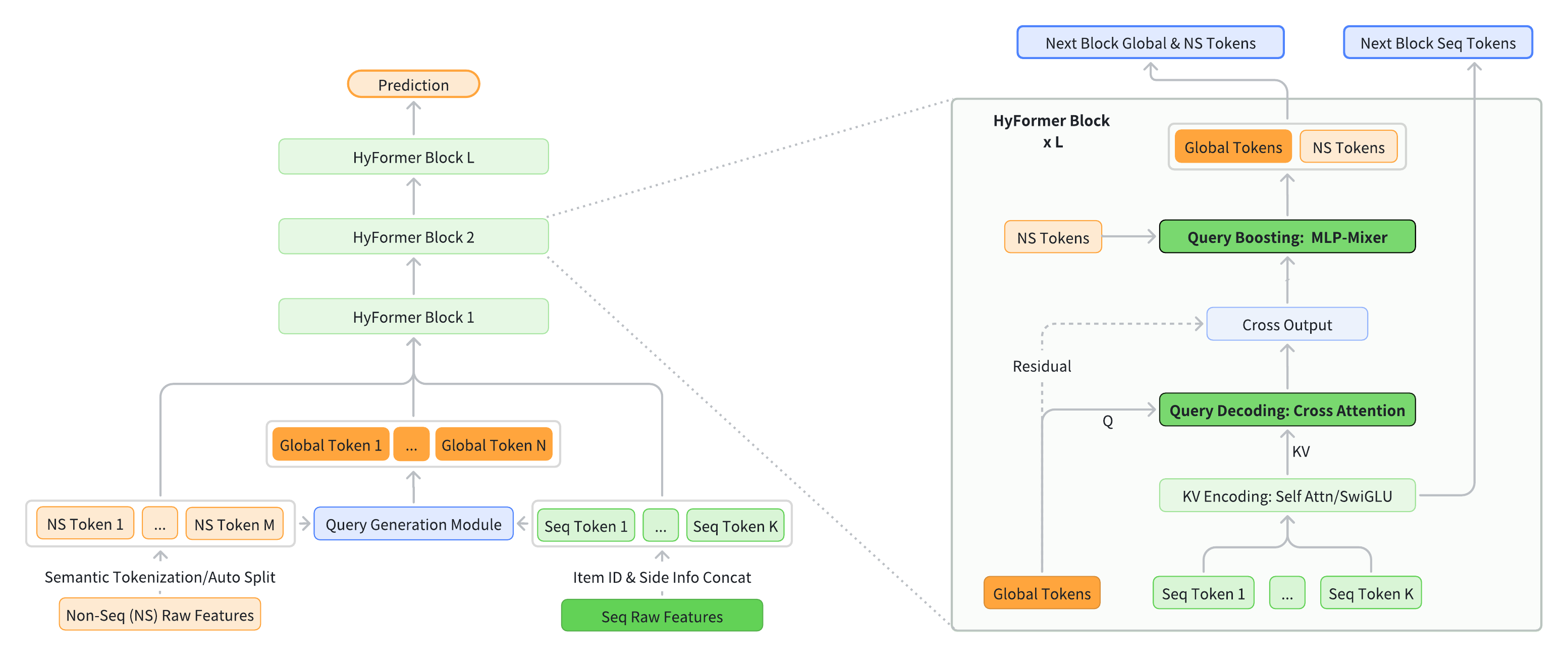}
    \caption{Overview of the proposed \textbf{HyFormer} architecture. The new arch introduces global tokens that derived from original "candidate item" in sequence modeling, and revisits the roles of long-sequence modeling and feature interaction by boosting the query capacity of long-sequence via MLP-Mixer-based feature interaction. It frames the LRM modeling task as an alternating optimization process through the alternation of \emph{Query Decoding} and \emph{Query Boosting} modules.}
    \label{fig:HyFormer}
\end{figure*}

\section{Methodology}


\subsection{Problem Statement}

Let $\mathcal{U}$ and $\mathcal{I}$ denote the user and item spaces. For a user
$u \in \mathcal{U}$, denote the raw behavioral history as
$S = [i^{(u)}_1, \ldots, i^{(u)}_K]$ with each
$i^{(u)}_t \in \mathcal{I}$, and let $u$ represent the accompanying
non-sequential descriptors such as profile attributes, contextual signals, and
cross features. Given a candidate item $v \in \mathcal{I}$, the goal is to
estimate the probability that user $u$ engages with item $v$:
\begin{equation}
P(y = 1 \mid S, u, v) \in [0,1],
\end{equation}
where $y \in \{0,1\}$ indicates whether the interaction occurs.

The model parameters are learned from historical data
$\mathcal{D} = \{(S, u, v, y)\}$ by minimizing the standard binary
cross-entropy objective:
\begin{equation}
\mathcal{L} = -\frac{1}{|\mathcal{D}|}
\sum_{(S, u, v, y) \in \mathcal{D}}
\Big[
    y \log \hat{y}
    +
    (1-y) \log (1 - \hat{y})
\Big],
\end{equation}
where $\hat{y} = f_{\theta}(S, u, v)$ denotes the predicted engagement
probability produced by the LRM.
    
    
\subsection{Overall Framework}



Traditional LRM architectures generally adopt a pipelined design by performing sequence modeling like LONGER \cite{chai2025longerscalinglongsequence} first, and the query token containing the compressed sequence information is then used for subsequence feature interaction like RankMixer\cite{zhu2025rankmixerscalingrankingmodels}. As discussed before, this separate pipeline generally results in an insufficient modeling for both sequence modeling and heterogeneous feature interaction. To overcome the limitation, this work proposes a unified hybrid framework that jointly models non-sequential (NS) tokens and long behavioral sequences through a stack of HyFormer layers.

The overall architecture of HyFormer is presented in Figure~\ref{fig:HyFormer}. As shown in the figure, each HyFormer layer integrates two complementary mechanisms:
(1) \emph{Query Decoding}, which expands non-sequential and sequential features into multiple semantic \textbf{global tokens} (i.e., sequence queries) via MLP-based query generation and performs cross-attention over long-sequence K/V pairs, enabling global information to directly shape the representation of sequence tokens; and
(2) \emph{Query Boosting}, which applies MLP-Mixer-style token mixing to strengthen interactions among decoded queries and non-sequence tokens. 


By tightly coupling global heterogeneous-feature mixing with efficient long-sequence modeling, the proposed framework achieves richer heterogeneous interactions, deeper utilization of sequential structure, and more favorable performance and computation cost compared with existing separate pipelined architectures.

\subsection{Query Generation }
\subsubsection{Input Tokenization}
Following the tokenization strategy in RankMixer\cite{zhu2025rankmixerscalingrankingmodels}, input tokens can be organized
either by \emph{semantic grouping} or by \emph{automatic splitting}.
Semantic grouping partitions tokens according to their intrinsic meanings
(e.g., user, context, or behavior semantics), while auto-split flattens all
features into a single embedding and applies uniform splitting without explicit
semantic distinctions. In practice, given the clear semantic roles of input
features in our setting, HyFormer adopts semantic grouping to preserve
structured inductive bias and improve interpretability.

\subsubsection{Query Generation}

The Query Generation module converts heterogeneous non-sequential features into
semantic query tokens used for decoding long behavioral sequences. All
non-sequential feature vectors $F_1, F_2, \ldots, F_M \in \mathbb{R}^{1 \times D}$ are concatenated and
mapped through a lightweight feed-forward network. In addition, a global
sequence-level summary is obtained via pooling over the behavioral sequence
representations and treated as an additional shared input, analogous to
non-sequential features.

The queries are generated by combining non-sequential features with the pooled
sequence summary through a lightweight projection:
\begin{equation}
Q =
\big[
\mathrm{FFN}_1(\mathrm{Global \ Info}), \ldots, \mathrm{FFN}_N(\mathrm{Global \ Info})
\big]
\in \mathbb{R}^{N \times D},
\end{equation}
where
\begin{equation}
\mathrm{Global \ Info} = \mathrm{Concat}\big(
F_{1}, \ldots, F_{M},\;
\mathrm{MeanPool}(Seq)
\big).
\end{equation}

To maintain serving efficiency, the module supports feature selection and
optional query compression, keeping the number of generated queries stable
while preserving sufficient representational capacity for downstream decoding.

In deeper HyFormer layers, queries are not regenerated through MLPs. Instead,
each layer reuses the queries from the previous layer, effectively using deeper
cross-attention outputs as updated queries to interrogate the long sequence
with progressively richer semantics.

\subsection{Query Decoding }
The Query Decoding module is responsible for transforming non-sequential features into semantic queries and extracting target-aware information from long behavioral sequences through cross attention. With the layer-wise key–value representations of the long sequence produced by the Sequence Representation Encoding module, Query Decoding module decodes the K/V representation with the multiple query tokens from the Query Generation Module via the multi-query cross attention mechanism.


\subsubsection{Sequence Representation Encoding}
HyFormer supports multiple sequence encoding strategies with different
capacity-efficiency trade-off. Given the behavioral sequence
$S$, each strategy produces layer-wise key--value representations
$(K^{(s)}_{l}, V^{(s)}_{l})$ for subsequent decoding.

\emph{(i) Full Transformer Encoding\cite{vaswani2017attention}.}
At the highest modeling capacity, a standard Transformer encoder is applied:
\begin{equation}
H_{l} = \mathrm{TransformerEnc}_{l}\big(S\big),
\end{equation}
which captures fine-grained interactions and long-range dependencies via full
self-attention.

\emph{(ii) LONGER \cite{chai2025longerscalinglongsequence}-style Efficient Encoding.}
To improve efficiency for long sequences, full self-attention is replaced by
cross-attention between a compact short sequence and the full history:
\begin{equation}
H_{l} =
\mathrm{CrossAttn}\big(
S_{\text{short}},\; S,\; S
\big),
\end{equation}
where $S_{\text{short}}$ is a compact short sequence with length $L_H \ll L_S$.
Here, $S_{\text{short}}$ serves as the query, while $S$ is used as both keys and
values. This formulation replaces full self-attention and reduces the
computational complexity from $\mathcal{O}(L_S^2)$ to
$\mathcal{O}(L_H L_S)$.

\emph{(iii) Decoder-style Lightweight Encoding.}
For latency-critical scenarios, sequence representations are transformed using
attention-free feed-forward operations:
\begin{equation}
H_{l} = \mathrm{SwiGLU}_{l}\big(S\big),
\end{equation}
trading contextual capacity for minimal computational cost.

Across all variants, the resulting representations are linearly projected to
obtain layer-specific key--value states:
\begin{equation}
K_{l} = H_{l} W^{K}_{l}, \qquad
V_{l} = H_{l} W^{V}_{l}.
\end{equation}
Key--value states are recomputed at each layer, allowing sequence features to
evolve jointly with decoder depth while supporting flexible deployment
configurations.

\subsubsection{Query Decoding via Cross-Attention}

Given the sequence-specific query tokens and the corresponding layer-wise
key--value representations, HyFormer performs Query Decoding through
cross-attention. For each behavioral sequence $S$ at layer $l$, the decoded
query representations are obtained as:
\begin{equation}
\tilde{Q}_{(l)} = \mathrm{CrossAttn}\!\left(
Q_{(l)},\,
K_{(l)},\,
V_{(l)}
\right),
\end{equation}
where $\mathrm{CrossAttn}(\cdot)$ denotes a standard multi-head cross-attention
operation, and $Q_{(l)} \in \mathbb{R}^{N \times D}$ represents the query token used at layer $l$.

This decoding step allows global, non-sequential features to directly attend to
long behavioral sequences, injecting contextual signals into the sequence-aware
query representations. The decoded query $\tilde{Q}_{(l)}$ are then used
as the semantic interface for subsequent interaction and boosting modules.

\subsection{Query Boosting}

The Query Boosting module enhances query representations before they are fed into
the subsequent cross-attention layer. After the decoding step, the queries
already encode sequence-aware information, but their interactions with static non-sequential heterogeneous features remain underexplored. Query Boosting
addresses this limitation by explicitly mixing information across query tokens
and injecting additional non-sequence-feature signals.

With the decoded output, the unified query representation is defined as
\begin{equation}
    Q = [\tilde{Q}_{(l)}, F_1, \ldots, F_M] \in \mathbb{R}^{T \times D},
\end{equation}
where $T = N + M$, $\tilde{Q}_{(l)} \in \mathbb{R}^{N \times D}$ denotes the set of
decoded query tokens obtained at layer $l$, and the remaining $M$ tokens
correspond to non-sequential feature embeddings.

Specifically, the boosting module applies an MLP-Mixer-style \cite{tolstikhin2021mlp} lightweight token-mixing operation inspired by RankMixer \cite{zhu2025rankmixerscalingrankingmodels} to enrich the decoded queries. Each query token $q_t \in Q$ is first partitioned into $T$ channel
subspaces:
\begin{equation}
q_t = [\, q_t^{(1)} \| q_t^{(2)} \| \cdots \| q_t^{(T)} \,],
\quad q_t^{(h)} \in \mathbb{R}^{D/T}.
\end{equation}

For each subspace index $h \in \{1,\ldots,T\}$, MLP-Mixer aggregates information
from all token positions by concatenating the corresponding subspaces:
\begin{equation}
\tilde{q}_h = \mathrm{Concat}\big(q_1^{(h)}, q_2^{(h)}, \ldots, q_T^{(h)}\big)
\in \mathbb{R}^{D}.
\end{equation}

Collecting all mixed tokens yields the token-mixed representation
\begin{equation}
\hat{Q} = [\tilde{q}_1, \tilde{q}_2, \ldots, \tilde{q}_T] \in \mathbb{R}^{T \times D}.
\end{equation}

The mixed queries are further refined by a lightweight per-token feed-forward
module:
\begin{equation}
\widetilde{Q} = \mathrm{PerToken\text{-}FFN}(\hat{Q}),
\end{equation}
where $\mathrm{PerToken\text{-}FFN}(\cdot)$ applies an independent feed-forward
transformation to each query token, enabling subspace-specific refinement while
preserving linear computational complexity.

Finally, a residual connection is applied to stabilize optimization and preserve
the original decoded semantics:
\begin{equation}
Q_{\mathrm{boost}} = Q + \widetilde{Q}.
\end{equation}

The boosted queries are then passed to the next HyFormer layer, allowing deeper
layers to interrogate long behavioral sequences with progressively richer and
more expressive representations.

\subsection{HyFormer Module}

The HyFormer module is constructed by stacking multiple layers, each consisting
of a \emph{Query Decoding} block followed by a \emph{Query Boosting} block.
At each layer, semantic queries interact with the long behavioral sequence via
cross-attention, and the resulting sequence-aware representations are further
refined to serve as inputs to deeper layers.

Formally, at layer $l$, the Query Decoding block takes the incoming global
queries $Q^{(l-1)}$ and performs cross-attention over the layer-wise key--value
representations $(K^{(l)}, V^{(l)})$ derived from the long sequence:
\begin{equation}
\widehat{Q}^{(l)} =
\mathrm{CrossAttn}\big(Q^{(l-1)}, K^{(l)}, V^{(l)}\big).
\end{equation}

The decoded queries $\widehat{Q}^{(l)}$ are then concatenated with the
non-sequential tokens and passed to the Query Boosting block, which applies
a lightweight token-wise transformation to enrich the query representations:
\begin{equation}
\widetilde{Q}^{(l)} =
\mathrm{QueryBoost}\big(
\mathrm{Concat}(\widehat{Q}^{(l)}, \mathrm{NS \ Tokens})
\big),
\end{equation}

By stacking multiple such layers, HyFormer progressively refines the semantic
queries, enabling deeper layers to abstract the long sequence with
increasingly expressive representations. The output of the top HyFormer layer is fed into downstream MLPs for final predictions, enabling efficient and flexible integration of heterogeneous non-sequential features with long behavioral sequences in LRMs.


\subsection{Multi-Sequence Modeling}

\begin{figure}[H]
  \centering
  \begin{minipage}{\linewidth} 
    \centering
    \includegraphics[width=\linewidth]{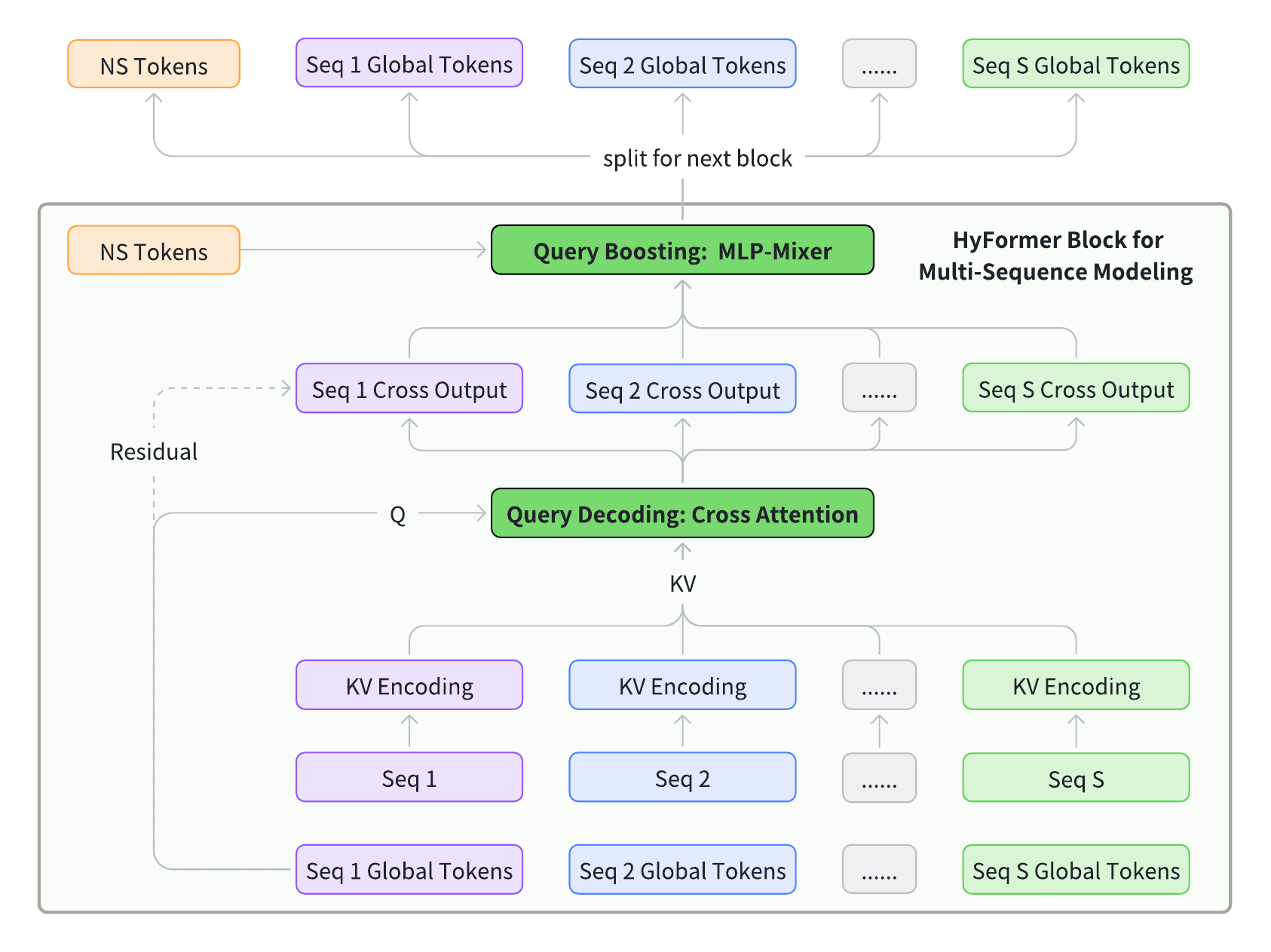}
    \caption{Multi-Sequence Modeling in HyFormer.}
    \label{fig:multi-seq}
  \end{minipage}
\end{figure}

In industrial recommendation scenarios, user behaviors are often organized as multiple heterogeneous sequences, for example, video-watch sequence and product-purchase sequence. As practical multi-sequences are generally with different feature spaces and semantic representations, we empirically find that simple sequence-merge that adopted by MTGR\cite{Han_2025} or OneTrans\cite{zhang2025onetransunifiedfeatureinteraction} would lead to a significant degradation in performance (see Section 4). Thus, instead of merging different sequences into a single unified stream, HyFormer processes each behavior sequence independently in each HyFormer block for both efficiency and effectiveness. As shown in Figure~\ref{fig:multi-seq}, for each sequence, a dedicated set of query tokens is constructed and used to perform Query Decoding over the corresponding sequence representations. This design preserves sequence-specific semantics during decoding, while enabling cross-sequence interaction to be handled later through query-level token mixing, without requiring explicit sequence concatenation.

\subsection{Training and Deployment Optimization}




\subsubsection{GPU Pooling for Long-Sequence.} User long‑sequence features can be extremely large, incurring significant data‑transfer overhead (e.g., Host‑to‑Device memory copies) and high memory pressure on the host. Fortunately, the number of truly unique feature IDs in such sequences is limited (typically ~25\% of the total tokens). We exploit this sparsity to deduplicate features, substantially reducing transfer costs and host‑memory footprint. Specifically, before graph execution, features are stored in a compressed embedding‑table. During execution, we build a high‑performance forward operator that reconstructs the original sequence features directly on the GPU. In the backward pass, the companion backward operator aggregates gradients from the sequence features into gradients for the embedding table. These gradients are then propagated upstream to update the sparse parameters.

\subsubsection{Asynchronous AllReduce}
To mitigate idle time introduced by synchronous gradient aggregation, the system enables asynchronous AllReduce, allowing the gradient synchronization of step $k$ to overlap with the forward and backward computation of step $k{+}1$. This design effectively eliminates communication bubbles and maximizes GPU utilization. The trade-off, however, is the introduction of one-step staleness for dense parameters: since their gradients are only available after the asynchronous reduction completes, the update rule becomes $W_{k} = W_{k-1} + g_{k-1}$
indicating that dense parameters at step $k$ use gradients from the previous step. In contrast, sparse parameters can be updated immediately after their local gradients are computed, yielding $W_{k} = W_{k-1} + g_{k}$
and thus effectively staying one step ahead of the dense parameter updates. Although this hybrid update schedule introduces a small degree of temporal inconsistency between dense and sparse parameter states, empirical results indicate that this staleness does not degrade convergence quality or model performance in practice.

\section{Experiments}

\begin{table*}[ht]
    \centering
    \setlength{\tabcolsep}{6pt}
    \caption{Overall Performance on Industrial Dataset}
    \label{tab:performance}
    \begin{tabular*}{\textwidth}{@{\extracolsep{\fill}}l l c c c c}
        \toprule 
        \textbf{Sequence Modeling} & \textbf{Feature Interaction} &
        \textbf{AUC$\uparrow$} & \textbf{$\Delta$AUC } &
        \textbf{\makecell{Params\\($\times10^{6}$)}} &
        \textbf{\makecell{FLOPs\\($\times10^{12}$)}} \\
        \midrule

        \multicolumn{6}{c}{\textbf{BaseArch: Traditional Two-Stage Models}} \\
        \midrule
        LONGER\cite{chai2025longerscalinglongsequence}
            & RankMixer\cite{tolstikhin2021mlp, zhu2025rankmixerscalingrankingmodels}
            & 0.6478 & --      & 386 & 3.5 \\
        LONGER\cite{chai2025longerscalinglongsequence}
            & Full Transformer\cite{vaswani2017attention}
            & 0.6472 & -0.09\% & 416 & 6.2 \\
        LONGER\cite{chai2025longerscalinglongsequence}
            & Wukong\cite{zhang2024wukong}
            & 0.6465 & -0.20\% & 385 & 5.2 \\
        Full Transformer\cite{vaswani2017attention}
            & RankMixer\cite{tolstikhin2021mlp, zhu2025rankmixerscalingrankingmodels}
            & 0.6481 & +0.05\% & 388 & 6.6 \\
        Full Transformer\cite{vaswani2017attention}
            & Full Transformer\cite{vaswani2017attention}
            & 0.6474 & -0.06\% & 418 & 9.3 \\
        Full Transformer\cite{vaswani2017attention}
            & Wukong\cite{zhang2024wukong}
            & 0.6468 & -0.15\% & 387 & 8.3 \\

        \midrule\midrule

        \multicolumn{6}{c}{\textbf{UniArch: Unified-Block Models}} \\
        \midrule
        \multicolumn{2}{l}{MTGR/OneTrans (w/ LONGER)\cite{Han_2025, zhang2025onetransunifiedfeatureinteraction}}
            & 0.6480 & +0.03\% & 406 & 6.6 \\
        \multicolumn{2}{l}{MTGR/OneTrans (w/ Full Transformer)\cite{Han_2025, zhang2025onetransunifiedfeatureinteraction}}
            & 0.6483 & +0.08\% & 450 & 21.9 \\
        \multicolumn{2}{l}{\textbf{HyFormer (Ours)}}
            & \textbf{0.6489} & +0.17\% & 418 & 3.9 \\

        \bottomrule
    \end{tabular*}
\end{table*}

\subsection{Experimental Setting}
\subsubsection{Datasets}
We evaluate our model on the Click Rate (CTR) prediction task in the Douyin Search System, a real-world and large-scale industrial search recommendation scenario at ByteDance. The dataset is derived from a subset of online user interaction logs spanning 70 consecutive days and comprises 3 billion samples. Each sample incorporates user features, query features, document features, cross-features, and several sequential features. The three primary sequences used in the model are defined as follows:
\begin{itemize}
\item Long-term sequence: The user's long-term search and click behavior sequence, the length of which can be tailored as needed, with an upper bound of 3000 adopted in this study.
\item Search sequence: The user's top-50 search behavior items, filtered by the Query Search module.
\item Feed sequence: The user's top-50 feed behavior items, filtered by the Query Search module.
\end{itemize}

\subsubsection{Baselines}
We compared our model against several strong baselines, which can be categorized into two architectural paradigms: Traditional Two-Stage Models and Unified-Architecture Models.

Traditional Two-Stage Models follow the prevalent mainstream design where sequence modeling and feature interaction are separated into two sequential stages. Specifically, sequential representations are first generated through a dedicated sequence modeling module and subsequently crossed with token-level representations of other features. For long-sequence modeling, we used the LONGER\cite{chai2025longerscalinglongsequence} or Full Transformer\cite{vaswani2017attention} architecture. To capture interactions between tokenized features, we employed several established architectures designed for feature interaction, including RankMixer\cite{zhu2025rankmixerscalingrankingmodels}, Full Transformer\cite{vaswani2017attention}, and Wukong\cite{zhang2024wukong}.

Unified-Block Models, in contrast, adopt a joint modeling approach where both sequential and non-sequential features are tokenized and processed simultaneously within a single model block. This integrates sequence modeling and heterogeneous feature interaction into one unified stage. An example is MTGR\cite{Han_2025}, which tokenizes all features and models them jointly using a transformer-style backbone. Similarly, OneTrans\cite{zhang2025onetransunifiedfeatureinteraction} follows a comparable simplified design, as it adopts a pyramid-compressed structure as backbone. In our implementation of the MTGR/OneTrans models, we performs MTGR/OneTrans (LONGER) for only cross‑attention between non‑sequential and sequential features, without the calculation of inner-sequence self-attention. Besides, we perform MTGR/OneTrans (Full Transformer) with full self-attention in sequence to achieve better performance with increased FLOPs.




\subsubsection{Evaluation Metrics}
For offline evaluation, we employ Query-level AUC (Area Under the Curve) which calculates the AUC\cite{hand2001simple} for samples within each query and then averages the results across all queries.  We also report the number of dense parameters and training FLOPs, with the latter computed using a batch size of 2048. 

\subsubsection{Implementation Details}
For the convenience of our experiment, the recommendation model is cold-started for offline evaluation and warmed up with checkpoints for online evaluation. All baselines utilize the same 2048 batch size and optimizer settings. The input token count for all MLPMixer modules is aligned to 16. In the multi-sequence HyFormer implementation, it comprises 13 non‑sequential tokens and 3 global tokens—one per sequence—summing to a total of 16 tokens. All models are trained with the same hyperparameter tuning, and experiments are conducted on a 64-GPUs cluster.

\subsection{Overall Performance}

\subsubsection{Comparison of existing methods.}
Our proposed HyFormer architecture achieves the highest AUC among all evaluated models, outperforming both traditional two-stage models (termed BaseArch) and other unified-block models (termed UniArch). Within the BaseArch group, performance varies significantly with component choice: for feature interaction, RankMixer~\cite{zhu2025rankmixerscalingrankingmodels} consistently outperforms Self-Attention and Wukong\cite{zhang2024wukong}, while for sequence modeling, incorporating full self-attention within the sequence generally yields gains. Notably, the best-performing BaseArch combination that employs a Full Transformer for sequence modeling with RankMixer still falls short of HyFormer, owing to its inherent limitation of unidirectional information flow. Furthermore, it is evident from the table that HyFormer demonstrates superior computational efficiency. Despite achieving the highest accuracy, it requires only 3.9×10¹² total FLOPs, including forward and backward propagation during training. This computational cost is significantly lower than that of most competitors, including other high-performing models such as MTGR\cite{Han_2025}. The overall performance results highlight the inherent limitations of the traditional two-stage paradigm. 

Unified architectures like HyFormer and MTGR demonstrate that integrating sequence modeling and feature interaction into a cohesive design enhances overall effectiveness. However, as evidenced by the results in the table, MTGR/OneTrans\cite{Han_2025,zhang2025onetransunifiedfeatureinteraction} relies on Self-Attention for feature interaction—an approach that often degrades AUC and significantly compromises computational efficiency in the interaction module\cite{zhu2025rankmixerscalingrankingmodels}. HyFormer, therefore, distinguishes itself by achieving the best accuracy without resorting to such costly substitutions or complex modeling on the sequence key-value side. This validates its core design principle of iterative query decoding and boosting within a unified block. Besides, MTGR/OneTrans combines Global Tokens and Seq Tokens as keys, while exclusively employing Global Tokens as queries. This design facilitates Global Tokens to attend more readily to themselves rather than sequence tokens. In contrast, HyFormer enforces a separated information flow: it first compresses and absorbs concrete sequence item information into Global Tokens, and then conducts interaction between different abstract Global Tokens, with this two-step process repeatedly stacked across layers. Furthermore, the hybrid architecture of HyFormer offers greater flexibility for future scaling. It allows for independent adjustment of interaction layers/dimensions and sequence modeling layers/dimensions, providing a more adaptable framework than methods that rigidly bind feature interaction and sequence modeling within a single standard attention layer. 

\subsubsection{Ablation Study.} Table~\ref{tab:ablation_components} presents an ablation study on the primary contributors to HyFormer's performance improvement. First, we ablate the components of the query. The HyFormer query is generated from three sources: global  non-sequential features, multiple sequences pooling tokens, and the original target features. Experiments show that reverting the query to its original, target-feature-only state severely limits subsequent deep feature interaction, causing a 0.08\% AUC decline. Removing the cross-sequence pooling tokens from the full query also leads to a 0.05\% AUC loss, confirming that inter-sequence interaction contributes meaningfully within the HyFormer structure.

Second, we evaluate the overall architectural change. Restoring the baseline architecture (LONGER + RankMixer), which applies sequential modeling followed by separate feature interaction, shows that even with enriched query information, the lack of deepened interaction caps the gains, yielding only a 0.03\% AUC improvement(-0.14\% vs -0.17\%). In contrast, within the HyFormer framework which is designed to strengthen interaction throughout the model, expanding query information delivers a significantly larger 0.08\% AUC gain.

\begin{table}[ht]
\footnotesize
\setlength{\tabcolsep}{3pt} 
\renewcommand{\arraystretch}{1.25}
\caption{Ablation Study on HyFormer Components}
\label{tab:ablation_components}
\centering
\begin{tabular}{lccccc} 
\toprule
\textbf{Configuration} & \textbf{AUC$\uparrow$} & \textbf{$\Delta$AUC} & \textbf{\makecell{Params\\($\times10^{6}$)}} & \textbf{\makecell{FLOPs\\($\times10^{12}$)}} \\
\midrule
\multicolumn{5}{l}{\textbf{Ablation of Query Global Context}} \\
\midrule
HyFormer                                  & 0.6489 & -       & 418 & 3.9 \\
Query w/o Seq Pooling Tokens              & 0.6486 & -0.05\% & 415 & 3.9 \\
Query w/o Nonseq and Seq Pooling Tokens   & 0.6484 & -0.08\% & 414 & 3.8 \\
\midrule
\multicolumn{5}{l}{\textbf{Ablation of Query Boosting}} \\ 
\midrule
HyFormer                   & 0.6489 & -       & 418 & 3.9 \\
HyFormer w/o Global Tokens & 0.6484 & -0.08\% & 414 & 3.8 \\
BaseArch w/ Global Tokens  & 0.6480 & -0.14\% & 505 & 3.6 \\
BaseArch w/o Global Tokens & 0.6478 & -0.17\% & 387 & 3.5 \\
\midrule
\multicolumn{5}{l}{\textbf{Ablation of Multi-Sequence Modeling}} \\ 
\midrule
HyFormer             & 0.6489 & -       & 418 & 3.9 \\
HyFormer + Merge Seq & 0.6485 & -0.06\% & 397 & 3.9 \\

\bottomrule
\end{tabular}
\end{table}

Third, we conduct an ablation study on the multi-sequence modeling strategy within HyFormer. Two principal paradigms exist for handling multiple sequences: merging sequences into one via dimension alignment and concatenation for joint modeling, or keeping sequences separate and modeling them independently. HyFormer adopts the latter approach, using distinct query tokens for each sequence. In our experiments, it is observed that the sequence merging and query sharing resulted in a significant AUC loss of 0.06\%. Consequently, this presents the advantages of the HyFormer in expanding queries and enabling broader feature interaction. Additionally, merging forces disparate sequences to share global tokens, ignoring their distinctiveness. The resulting representations capture far less differentiated information than HyFormer's separate modeling of each sequence. We speculate that this inherent limitation of sequence merging also partly explains why models like MTGR and OneTrans underperform compared to HyFormer.

In summary, the HyFormer architecture provides a versatile multi-sequence modeling framework by employing independent tokens for different sequences, thereby eliminating the need for forced alignment of side information or sparse dimensions across sequences. This design not only preserves the inherent distinctions between sequences to a great extent, but also enables the adaptive allocation of more global tokens to more important sequences, which has yielded measurable gains in our offline experiments.


\subsection{Scaling Analysis}

In this section, we present the scaling analysis of model performance with respect to sequence side information, FLOPs, and the number of parameters. 
As shown in the overall performance in Table~\ref{tab:performance}, under the paradigm of first performing sequential modeling and then performing heterogeneous feature interaction, LONGER + RankMixer achieves the best performance and is currently the production baseline. Therefore, we use it as the control group as BaseArch in the scaling experiments to compare the scaling performance of the HyFormer architecture.

\subsubsection{Parameters \& FLOPs}

\begin{figure}[H]
  \centering
  \begin{subfigure}[b]{0.48\linewidth}
    \centering
    \includegraphics[width=\linewidth]{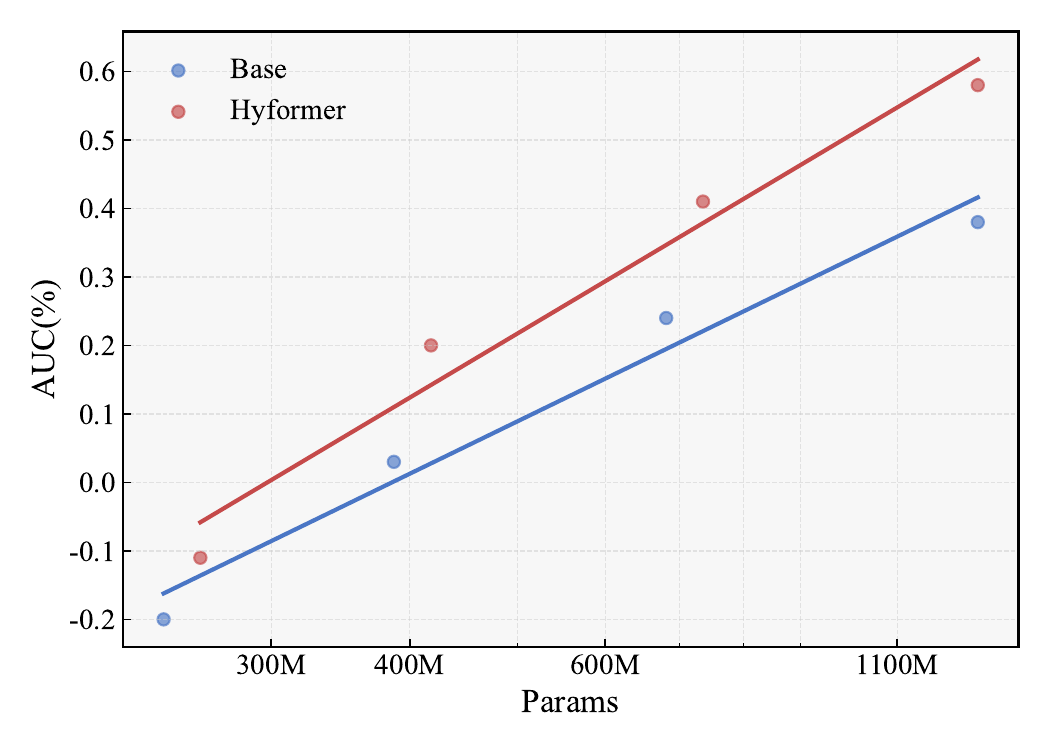}
    \caption{AUC Scaling with Params}
    \label{fig:parameter_qauc}
  \end{subfigure}
  \hfill
  \begin{subfigure}[b]{0.48\linewidth}
    \centering
    \includegraphics[width=\linewidth]{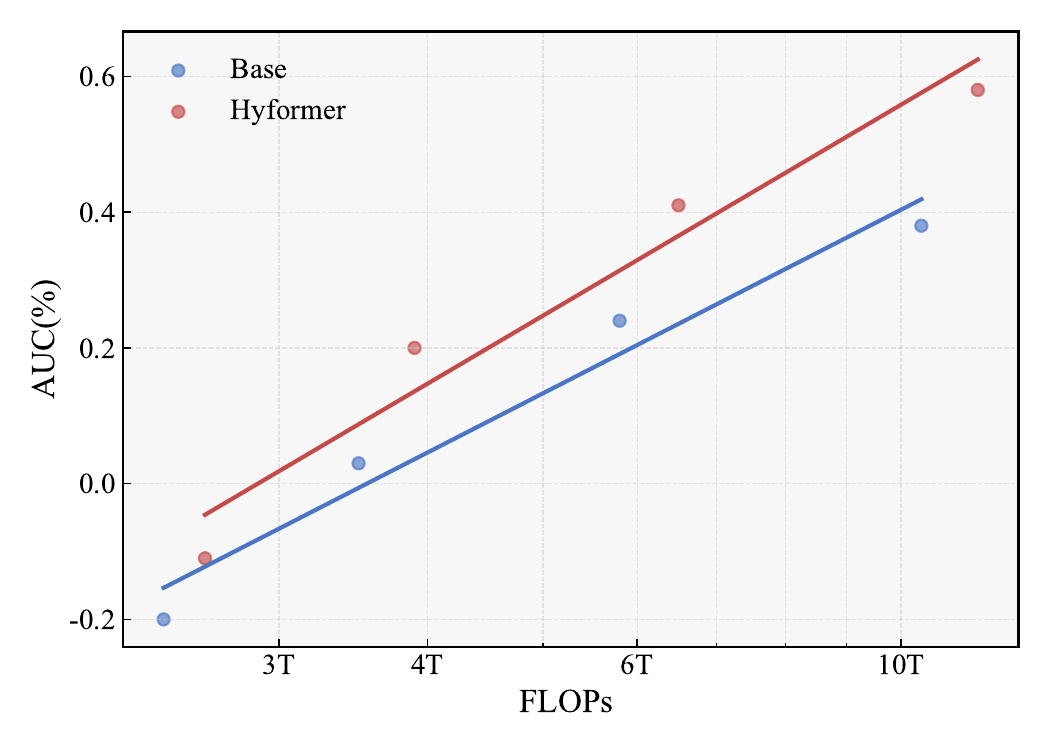}
    \caption{AUC Scaling with FLOPs }
    \label{fig:flops_qauc}
  \end{subfigure}
  \caption{Scaling performance with respect to FLOPs and model parameters.}
  \label{fig:scaling_flops_params}
\end{figure}

We examine the scaling law of the HyFormer architecture across model sizes ranging from 200M to 1B+ parameters. The results are shown in Figure~\ref{fig:scaling_flops_params}(a). As can be observed, while HyFormer initially outperforms the baseline LONGER + RankMixer model, it maintains strong scaling benefits overall, exhibiting a steeper slope than the baseline. This indicates that the bidirectional information flow, enabled by the alternating stacked layers of the LONGER and RankMixer in HyFormer, allows it to achieve significantly greater gains from increasing depth compared to the baseline at similar parameter scales. 
A similar pattern emerges when scaling law is analyzed in terms of computational cost (FLOPs). As shown in Figure~\ref{fig:scaling_flops_params}(b), AUC increases steadily with FLOPs, following a strong power-law trend. This indicates that increasing computational resources enables the model to process sequences with richer information, benefiting from the expansion of the initial query and the repeated enhancement of the query through feature interaction in MLP-Mixer, ultimately leading to greater AUC improvement.

These results suggest that the architectural design of HyFormer prioritizes scaling efficiency, yielding greater gains per parameter via enriched heterogeneous feature interactions, which results in a steeper performance scaling curve.

\subsubsection{Sparse Dim}

\begin{table}[ht]
\footnotesize	
\setlength{\tabcolsep}{5pt}  
\renewcommand{\arraystretch}{1.25}
\caption{Scaling with Sequence Sparse Dim}
\label{tab:sparse_dim}
\centering
\scalebox{0.92}{
\begin{tabular}{lcccccc}
\toprule
\textbf{Seq Length} & \textbf{Arch} & \textbf{Seq Sparse Dim} &  \textbf{AUC$\uparrow$} & \textbf{$\Delta$AUC} & \textbf{$\Delta$AUC Gap} \\
\midrule
\multirow{4}{*}{1k}
 & \multirow{2}{*}{BaseArch} & 64  & 0.6478 & -       & -       \\
 &                           & 224 & 0.6484 & +0.09\% & -       \\
 \cline{2-6}
 & \multirow{2}{*}{HyFormer} & 64  & 0.6489 & -       & -       \\
 &                           & 224 & 0.6497 & +0.12\% & +0.03\% \\
\midrule
\multirow{4}{*}{3k}
 & \multirow{2}{*}{BaseArch} & 64  & 0.6486 & -       & -       \\
 &                           & 224 & 0.6490 & +0.06\% & -       \\
 \cline{2-6}
 & \multirow{2}{*}{HyFormer} & 64  & 0.6499 & -       & -       \\
 &                           & 224 & 0.6507 & +0.12\% & +0.06\%  \\
\bottomrule
\end{tabular}}
\end{table}

We also analyzed how model performance varies with the expansion of the sequence token input dimension (sparse embedding dim), i.e., the richness of sequence side information. Our experiments show that, regardless of sequence length, enriching sequence side information consistently brings greater benefits to the HyFormer framework than to the baseline LONGER + RankMixer framework. As shown in Table~\ref{tab:sparse_dim}, for sequences of length 1000, expanding the sparse dimension width from the original 64 dimensions with three side information types (item ID, search query textnet classification, and timestamp) to 224 dimensions with seven types (adding search query ID, author ID, event ID, and playtime) yielded a $\Delta$AUC of 0.09\% for the baseline, compared to a 0.12\% gain for HyFormer. The improvement for HyFormer is significantly larger, a trend that holds across other sequence lengths in experiments. Furthermore, the performance gap between HyFormer and the BaseArch widens as sequences grow longer, with the additional gain from dimension expansion increasing from 0.03\% at 1k sequence length to 0.06\% at 3k.

These results indicate that expanding the sequence key/value information delivers greater value within the HyFormer framework, and this advantage becomes more pronounced with longer sequences. The benefit stems from HyFormer's ability to integrate richer global information into sequence queries, coupled with the bidirectional information flow between its LONGER and Mixer modules, which collectively enable more thorough feature interaction.

\subsection{Online A/B Tests}

This section presents the online A/B test results for HyFormer model on the Douyin Search platform, where it was evaluated against strong existing RankMixer baseline. For online evaluation, we employ three key metrics: Average Watch Time Per User, Video Finish Play Count Per User, and Query Change Rate. In particular, the Query Change Rate quantifies the probability of a user manually refining a search query to a more specific one (e.g., from “iPhone” to “iPhone 17 Pro”), which is calculated as follows:

\begin{equation}
    \begin{split}
\mathrm{query \ change \ rate} = \frac{N_{\mathrm{reform}}} {N_{\mathrm{total}}}
    \end{split}
\end{equation}
where $N_{\mathrm{reform}}$ is the number of distinct user-query pairs with query reformulation, and $N_{\mathrm{total}}$ is the total number of distinct user-query pairs. This metric serves as an indicator of a negative search experience for users.


As shown in Table~\ref{tab:ab_test_results}, the online A/B Test confirms substantial improvements across key metrics: a 0.293\% increase in average watch time per user, a 1.111\% growth in video finish play count per user and a 0.236\% decrease in query change rate. These significant gains demonstrate the practical value and effectiveness of HyFormer in a real‑world, billion‑user platform environment.

\begin{table}[htbp]
\centering
\caption{Online A/B Test Results on Douyin}
\label{tab:ab_test_results}
\begin{tabular}{lc}
\toprule
\textbf{Online Test Metrics} & \textbf{Gain} \\
\midrule
Average Watch Time Per User $\uparrow$          & +0.293\% \\
Video Finish Play Count Per User $\uparrow$     & +1.111\% \\
Query Change Rate $\downarrow$                  & -0.236\% \\
\bottomrule
\end{tabular}
\end{table}

\section{Conclusions}

In this paper, we propose the HyFormer architecture. In contrast to the prevalent "Long Sequence Modeling, Then Feature Interaction" paradigm which first performs sequential modeling and then conducts heterogeneous feature interaction in a unidirectional flow, HyFormer introduces Global Tokens to redefine the roles of long-sequence modeling and feature interaction by boosting the query capacity via feature interaction. The architecture alternates between two core components: Query Decoding and Query Boosting. From a sequential modeling perspective, this corresponds to an iterative optimization process that alternates between decoding long sequences using the Global Tokens and enhancing the Global Tokens through cross-feature interaction. This design provides a novel and effective framework for more thorough sequence modeling and feature interaction, while also providing a flexible paradigm for multi-sequence modeling. Extensive offline and online experiments validate the superiority of upgrading from a unidirectional information flow to a bidirectional, co-evolutionary paradigm, and also raise the scaling ceiling for future LRMs in industry.

\balance
\bibliographystyle{ACM-Reference-Format}
\bibliography{main}


\end{document}